\documentclass[sigconf,screen,9pt]{acmart}
\acmConference[EASE 2026]{The 30th International Conference on Evaluation and Assessment in Software Engineering}{9–12 June, 2026}{Glasgow, Scotland, United Kingdom}

\newif\ifEditMode

\EditModetrue

\usepackage{vu}
\usepackage{aliases}

\usepackage[acronym]{glossaries}

\makeglossaries

\newacronym{llm}{LLM}{Large Language Model}
\newacronym{ros}{ROS}{Robot Operating System}

\renewcommand*{\acrshort}[1]{\textcolor{black}{\glsentryshort{#1}}}
\renewcommand*{\acrlong}[1]{\textcolor{black}{\glsentrylong{#1}}}

\usepackage[most]{tcolorbox}

\newtcolorbox{answerbox}{
  colback=brightBackground,
  colframe=black!40,
  boxrule=0.5pt,
  arc=1pt,
  left=1pt,
  right=1pt,
  top=1pt,
  bottom=1pt,
  before skip=0.1\baselineskip,
  after skip=0.1\baselineskip
}

\definecolor{brightBackground}{RGB}{230, 230, 230}

\begin{document}










\title{Can Large Language Models Assist the Comprehension of\\ ROS2 Software Architectures?}

\author{Laura Duits}
\affiliation{
  \institution{Vrije Universiteit Amsterdam}
  \country{The Netherlands}
}
\email{l.b.m.duits@student.vu.nl}

\author{Bouazza El Moutaouakil}
\affiliation{
  \institution{Vrije Universiteit Amsterdam}
  \country{The Netherlands}
}
\email{b.elmoutaouakil@vu.nl}

\author{Ivano Malavolta}
\affiliation{
  \institution{Vrije Universiteit Amsterdam}
  \country{The Netherlands}
}
\email{i.malavolta@vu.nl}

\begin{abstract}

\noindent \textit{Context}. 
The most used development framework for robotics software is \acrshort{ros}2. ROS2 architectures are highly complex, with thousands of components communicating in a decentralized fashion.  

\noindent \textit{Goal}. 
We aim to evaluate how \acrshort{llm}s can assist in the comprehension of factual information about the architecture of \acrshort{ros}2 systems.

\noindent \textit{Method}. 
We conduct a controlled experiment where we administer 1,230 prompts to 9 \acrshort{llm}s containing architecturally-relevant questions about 3 \acrshort{ros}2 systems with incremental size. We provide a generic algorithm that systematically generates architecturally-relevant questions for a \acrshort{ros}2 system. 
Then, we (i) assess the accuracy of the answers of the LLMs against a ground truth established via running and monitoring the 3 \acrshort{ros}2 systems and (ii) qualitatively analyse the explanations provided by the LLMs. 

\noindent \textit{Results}. 
Almost all questions are answered correctly across all LLMs (mean=98.22\%). \texttt{gemini-2.5-pro} performs best (100\% accuracy across all prompts and systems), followed by \texttt{o3} (99.77\%), and \texttt{gemini-2.5-flash} (99.72\%); the least performing \acrshort{llm} is \texttt{gpt-4.1} (95\%). Only 300/1,230 prompts are incorrectly answered, of which 249 are about the most complex system. 
The coherence scores in LLM's explanations range from 0.394 for ``service references'' to 0.762 for ``communication path''. The mean perplexity varies significantly across models, with \texttt{chatgpt-4o} achieving the lowest score (19.6) and \texttt{o4-mini} the highest (103.6).

\noindent \textit{Conclusions}. 
There is great potential in the usage of \acrshort{llm}s to aid \acrshort{ros}2 developers in comprehending non-trivial aspects of the software architecture of their systems. Nevertheless, developers should be aware of the intrinsic limitations and different performances of the LLMs and take those into account when using them. 

\end{abstract}

\maketitle


\section{Introduction}\label{s:intro}


The \acrlong{ros} (\acrshort{ros}) can be considered as the de-facto standard for developing robotics systems~\cite{Macenski:2022}. \acrshort{ros} supports modular development and integration of third-party components~\cite{Quigley:2009, Macenski:2022}. In 2004, more than 531M ROS packages were downloaded ~\cite{openrobotics2024metrics}. \acrshort{ros}2 provides standardised communication mechanisms, including topics, services, and actions, allowing components to seamlessly exchange information. ROS nodes can be created and removed dynamically, allowing flexibility and reconfigurability. Additionally, \acrshort{ros} provides reusable libraries (\eg \texttt{tf2}, \texttt{rclcpp}) and tools (\eg \texttt{rviz}, \texttt{rqt\_graph}), which accelerate development and facilitate integration on a variety of robotic platforms~\cite{JSS_ROS_2021,JSS_2023_ROS_SE}. 

The capabilities of \acrlong{llm}s (\acrshort{llm}s) are advancing rapidly and are being adopted across multiple domains ~\cite{chkirbene2024large}. \acrshort{llm}s are already used for various software engineering tasks~\cite{Schmid:2025}, such as code generation~\cite{Wang:2023}, automated documentation~\cite{Diggs:2025}, test case generation~\cite{Zhang:2025}, and assisting software architects in design decision making~\cite{Dhar:2024,esposito2025generative}. However, \textit{the potential of LLMs to support the \textit{comprehension} of the software architecture of a robotic system remains largely unexplored}.
%
In this context, the ability of LLMs to process large amounts of data and internalise cross-domain knowledge makes them good candidates for comprehending the software architecture of complex systems like those based on \acrshort{ros}.
Before these models can be effectively used to aid \acrshort{ros} developers in tasks such as system design, verification, or optimisation, it is necessary to establish their \textit{comprehension} about \acrshort{ros} systems. 

Our \textbf{goal} is to investigate the ability of \acrshort{llm}s to aid architects in comprehending the software architecture of \acrshort{ros}2 robotics systems. To this end, we carry out a controlled experiment involving 9 commercial \acrshort{llm}s and 3 third-party \acrshort{ros}2 systems. By defining and integrating an automated question-generation algorithm with a structured prompt design methodology, a total of 29,406 prompts are systematically generated. We sample 1,230 questions, prompt them to each of the 9 LLMs, and compare the 11,070 responses of all \acrshort{llm}s to answer questions about the three different \acrshort{ros}2 systems. The responses are compared against a predefined ground truth and further analysed for general patterns, emerging topics, length, and perplexity score.

The \textbf{results} of our study show that \acrshort{llm}s can reach good accuracy in comprehension tasks about the architecture of \acrshort{ros}2 systems. However, there are aspects where the knowledge or interpretation of the models differs.
Specifically, the majority of questions were answered correctly across all LLMs (mean = 98.22\%), with even a 100\% accuracy on the simplest system.
\texttt{gemini-2.5-pro} performs best (100\% accuracy across all prompts and systems), followed by \texttt{o3} (99.77\%), and \texttt{gemini-2.5-flash} (99.72\%); the least performing \acrshort{llm} is \texttt{gpt-4.1} (95\%). Over the 11,070 prompts, only 300 were incorrectly answered, of which 249 are about the largest system. About 99\% (297/300) of the incorrect answers refer to one specific type of question, \ie the one about the presence of a message exchange between two ROS nodes.   
The coherence score of the main topics present in the explanations provided by the LLMs ranges from 0.394 for the ``service references'' topic to 0.762 for the ``communication path'' topic. There is a wide disparity in the mean perplexity of the LLMs with \texttt{chatgpt-4o} having the lowest score (19.6) and \texttt{o4-mini} having the highest (103.6).

The \textbf{main contributions} of this study are:
(i) an empirical evaluation of the ability of \acrshort{llm}s to assist architects in comprehension tasks about \acrshort{ros}2 systems; 
(ii) an algorithm for systematically generating questions about a \acrshort{ros}2 system topology;
(iii) an in-depth discussion and contextualization of the obtained results for both academics and \acrshort{ros} developers;
(iv) the replication package of the study including the source code  for extracting the topologies, generating questions, establishing the prompts, and prompting the \acrshort{llm}s, as well as the raw data, and data analysis scripts~\cite{replication_package}.


\section{Background}\label{s:background}


ROS~\cite{Quigley:2009} is a widely adopted framework for robotic software. Despite its popularity, \acrshort{ros}1 was limited in terms of security, reliability, and scalability. To address these concerns, \acrshort{ros}2 was redesigned from scratch~\cite{Macenski:2022}. Key architectural differences are the usage of the Data Distribution Service (DDS)~\cite{Castellote:2003} as middleware, allowing for more reliable and scalable communication, and the inclusion of DDS-Security~\cite{Beckman:2018}. 
\acrshort{ros}1 has reached end-of-life on 31 May 2025; consequently, this work focusses exclusively on \acrshort{ros}2, and in this context, ROS will be considered synonymous with \acrshort{ros}2.  

\acrshort{ros} systems comprise executable programs called \textbf{nodes}, each performing a specific task inside the system. Nodes can be organised into \textbf{packages} that can be distributed and reused. There are three ways for nodes to communicate with each other. \textbf{Topics} can be used for asynchronous messaging between publishers and subscribers. \textbf{Services}, on the other hand, enable synchronous communication in which a client sends a request to a service server and, upon completion, receives a response in return. Finally, it is also possible to use \textbf{actions} by having a client sending a goal to an action server, who sends feedback and, upon reaching the goal, the result. However, as of today, actions are less frequently used. 
To allow for the configuration of specific variables, \acrshort{ros} nodes have a \textbf{parameter} feature.  Parameters can be configured at system launch, but also updated at runtime by a user or other nodes. The parameter changes are node specific; however, this information might be relevant for other nodes as they might want to react to these changes as well. For this purpose, there is a special topic called \texttt{'/parameter\_events'} on which all parameter changes are published. 

\vspace{-2mm}
\begin{figure}[htbp!]
    \centering
    \includegraphics[width=\linewidth]{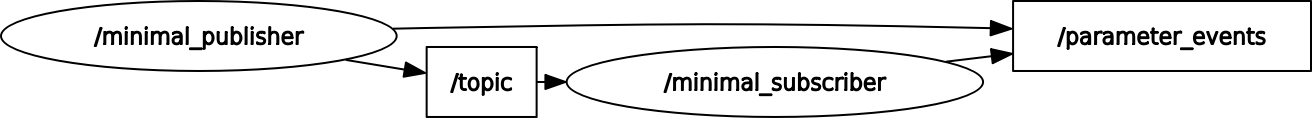}
    \vspace{-6mm}
    \caption{Computation graph of the \texttt{\textbf{pubsub}} system.}
    \label{fig:computation_graph}
\end{figure}
\vspace{-4mm}

A \textbf{computation graph} (or topology) illustrates the network of nodes and their communication interfaces within a \acrshort{ros} system. Figure~\ref{fig:computation_graph} presents an example of a computation graph for a simple system consisting of two nodes (\ie publisher and subscriber), a topic for their communication, and the system-wide \texttt{'/parameter\_events'} topic~\cite{PUBSUB}.
For the purpose of this study, we exclude action servers from the analysis, meaning that we exclusively consider topics and service servers as interfaces for the nodes. In addition, we consider the system-level topic \texttt{'/parameter\_events'} as a valid communication path that links two nodes with each other.

\section{Related Work}\label{s:related}


Various architectural tasks can be supported by \acrshort{llm}s~\cite{Schmid:2025}. 
Referring to \autoref{tab:related}, Dhar~\etal~\cite{Dhar:2024} investigated the feasibility of \acrshort{llm}s to generate design decisions for Architecture Design Records (ADRs). They compared the performance of two types of \acrshort{llm}s: GPT for decoder-only models and T5 for encoder-decoder models. 
The results indicate that \acrshort{llm}s achieve an overall good BERTScore on the 95 ADRs prompted (\eg GPT-4 shows a precision of 0.847 and an F1 of 0.849 in the 0-shot), indicating the potential to help architects document design decisions. Nevertheless, the authors report that the quality of the responses is insufficient, necessitating human involvement in the process.
Where \cite{Dhar:2024} used a single-shot, D{\'i}az-Pace~\etal~\cite{DiazPace:2024} consider a step-wise approach using five \acrshort{llm}-based copilots that allow user interaction to generate design decisions. In addition, the work included the usage of RAG strategies to improve the results. Nevertheless, the copilots struggled to formulate the decisions due to a lack of context. 

\acrshort{llm}s can also support architects in their understanding of a system. 
Soliman and Keim~\cite{Soliman:2025} evaluated the ability of GPT-3.5 to answer architectural knowledge questions about a software system embedded in the model. A zero-shot approach was used to prompt the questions. The results were compared against a ground truth to determine the model's accuracy and were manually assessed on the quality and trustworthiness. The responses were considered moderately accurate while often including irrelevant information. However, they did analyse the potential to help practitioners understand the architecture of existing software systems. 
This work was extended by \cite{Soliman:2026} to compare the performance of seven prominent \acrshort{llm} when answering the same set of AK questions. The findings align with previous work on the accuracies of the models. Moreover, while there are often similarities between the responses of some of the \acrshort{llm}s, these similarities do not extend to all seven models.

Amalfitano~\etal~\cite{Amalfitano:2026} assessed whether \acrshort{llm}s are able to automate software architecture recovery (SAR) based on source code. Their approach considered 3 types of tasks related to class diagrams, architectural and design patterns, and architectural styles. They considered 4 widely adopted \acrshort{llm}s: GPT-4o, Gemini 1.5 Pro, Claude 3.5 Sonnet, and Mistral Large. The study included a self-reflection mechanism to encourage the \acrshort{llm}s to evaluate the correctness of their  answers. Although the models were generally effective in identifying high-level architectural styles and basic structural elements, they struggled with implementation-specific analyses and less conventional architectural styles. Self-reflection improved performance on tasks involving structural and conceptual abstractions, but had little effect on fine-grained pattern recognition.


\begin{table*}[t]
    \centering
    \small
    \caption{Overview of the most recent related work on LLMs applied to software architecture tasks}
    \label{tab:related}
    \footnotesize
    \vspace{-3mm}
    \begin{tabular}{p{3.0cm} c p{1.7cm} p{4.2cm} p{2.8cm} p{4cm}}
        \hline
        \textbf{Study} & \textbf{Year} & \textbf{Target task} & \textbf{\acrshort{llm}s} & \textbf{Approach} & \textbf{Key findings \& Limitations} \\ \hline

    Amalfitano~\etal Automated Software Architecture Design Recovery from Source Code Using LLMs~\cite{Amalfitano:2026} & 2026
            & Automated recovery of architectural knowledge (AK) from source code
            & \texttt{GPT-4o, Gemini 1.5 Pro, Claude 3.5 Sonnet, Mistral Large}
            & 0-shot combined with role and emotion prompting followed by self-reflection of the responses compared to a pre-defined ground truth
            & LLMs partially recover architectures and styles but struggle with deeper abstractions and semantic consistency, partly improved by self-reflection; Comparison based on correctness
              \\ 

     \rowcolor{vu-grey-50}   Soliman~\etal LLMs for Software Architecture Knowledge: A Comparative Analysis Among Seven LLMs~\cite{Soliman:2026} & 2026
            & Answer questions about the AK of Apache Hadoop HDFS
            & \texttt{Qwen2-Instruct, DeepSeek-R1-Distill-Llama, GPT-4o, GPT-4o-mini, Llama-3.3-Instruct, TinyLlama-Chat-v1.0, Mistral-Large-Instruct-2407}
            & 0-shot responses compared to a pre-defined ground truth
            & There is a high number of false positives but some similarity between the \acrshort{llm}s although not uniform across all models; Usage of a single software system causing it to be less generalisable
              \\ 
        
         Soliman \& Keim. Do Large Language Models Contain Software Architectural Knowledge?~\cite{Soliman:2025} & 2025 
            & Answer questions about the AK of Apache Hadoop HDFS
            & \texttt{GPT-3.5} 
            & 0-shot responses compared to a pre-defined ground truth
            & Moderately accurate answers, often containing irrelevant information; Usage of a single software system and \acrshort{llm} causing it to be less generalisable
              \\ 
        

     \rowcolor{vu-grey-50}   Dhar~\etal Can LLMs Generate Architectural Design Decisions?~\cite{Dhar:2024} & 2024 
            & Generate Architecture Design Records (ADRs) 
            & \texttt{GPT2, GPT2-medium, GPT2-large, GPT2-xl, ada, davinci, text-davinci-003, GPT-3.5-turbo, GPT-4, T5-small, T5-base, T5-large, T5-3b, T0-3b, Flan-T5-small, Flan-T5-base, Flan-T5-large, Flan-T5-xl}
            & 0-/few-shot and fine-tuning comparisons for ADR Decision generation based on a given decision context
            & LLMs can produce plausible ADR decisions but quality is insufficient for autonomous use; Focus on text generation rather than reasoning correctness  \\

        Díaz-Pace~\etal Helping Novice Architects to Make Quality Design Decisions Using an LLM-Based Assistant~\cite{DiazPace:2024} &  2024 
            & LLM-based agent for architectural decision-making activities
            & \texttt{GPT-3.5}
            & 0-shot and Retrieval-Augmented Generation (RAG) for prompting architectural questions   
            & Usage of RAG improves results using a 0-shot approach but a lack of context forms a bottleneck; The \acrshort{llm} is integrated into a pipeline, which prevents its results from being analysed in isolation. 
              \\ \hline
        
    \end{tabular}
\end{table*}

In conclusion, to the best of our knowledge, our study is the first to evaluate the architectural knowledge of \acrshort{llm}s concerning \acrshort{ros}2 systems. To this end, we extend the findings of \cite{Soliman:2025} and \cite{Soliman:2026} by considering different and multiple software systems. Moreover, we consider both the answers and the explanations provided by the \acrshort{llm}s, providing insight into the reasoning capabilities of LLMs. 
\begin{table*}[htbp!]
\centering
\caption{Overview and main characteristics of the analysed \acrshort{ros}2 systems.}
\footnotesize
\vspace{-3mm}
\begin{tabular}{>{\centering\arraybackslash}m{1.3cm}|
                >{\centering\arraybackslash}m{1.5cm}
                >{\centering\arraybackslash}m{1cm}
                >{\centering\arraybackslash}m{5.3cm}|
                >{\centering\arraybackslash}r
                >{\centering\arraybackslash}r
                >{\centering\arraybackslash}r
                >{\centering\arraybackslash}r|
                >{\centering\arraybackslash}l}

        \textbf{System} & \textbf{Maintainer} & \textbf{Year} & \textbf{\acrshort{ros} Packages}  & \textbf{Nodes} & \textbf{Topics} & \textbf{Services} & \textbf{Connections} & \textbf{GitHub Repo} \\ \hline
        
        \texttt{\textbf{pubsub}} & - & - & \texttt{-} & 2 & 3 & 12 & 4 & \href{https://github.com/ros2/ros2_documentation/blob/humble/source/Tutorials/Beginner-Client-Libraries/Writing-A-Simple-Py-Publisher-And-Subscriber.rst}{\texttt{ros2\_documentation}} \\ 
        
        \rowcolor{vu-grey-50}\texttt{\textbf{turtlebot}} & ROBOTIS & 2018 & \shortstack{\texttt{gazebo}, \texttt{cartographer},  \texttt{navigation2}} & 8 & 15 & 53 & 64 & \href{https://github.com/ROBOTIS-GIT/turtlebot3?utm_source=chatgpt.com}{\texttt{turtlebot3}} \\ 
        
        \texttt{\textbf{panda}} & IFRA Group & 2022 & \shortstack{\texttt{gazebo}, \texttt{moveit}, \texttt{ros2\_control}, \texttt{ros2\_controllers},\\ \texttt{gripper\_controllers}, \texttt{xacro}} & 40 & 33 & 276 & 402 & \href{https://github.com/IFRA-Cranfield/ros2_RobotSimulation/tree/humble}{\texttt{ros2\_RobotSimulation}} \\ \hline
        
\end{tabular}
\label{tab:systemss}
\end{table*}

\section{Study Design}\label{s:design}



        
        


\subsection{Research Questions}
According to the template by Basili \etal~\cite{basili1988tame}, the \textbf{goal} of this study is to 
analyse \textit{the responses of \acrshort{llm}s}
for the purpose of \textit{evaluating} 
their \textit{output size, accuracy, perplexity, and mentioned topics} 
from the point of view of \textit{\acrshort{ros}2 developers} 
in the context of \textit{\acrshort{ros}2 software architectures}.
The research questions of this study are:

\noindent \textbf{$\mathbf{RQ_1}$ -- What is the \underline{output size} of \acrshort{llm}s when answering questions related to the ROS2 software architectures?} 
This RQ sets the stage and helps ROS developers understanding the verbosity and (possible) resource consumption of \acrshort{llm}s, which directly impact usability and cost in real-world ROS2 development scenarios.

\noindent \textbf{$\mathbf{RQ_2}$ -- What is the \underline{accuracy} of \acrshort{llm} when answering questions related to the ROS2 software architectures?}
Measuring accuracy is essential to determine whether LLMs can reliably provide correct architectural information, ensuring their practical applicability for supporting ROS2 system design and validation.

\noindent \textbf{$\mathbf{RQ_3}$ -- What is the level of \underline{perplexity} of \acrshort{llm}s when answering questions related to the ROS2 software architectures?}
This RQ complements $\mathbf{RQ_2}$ since perplexity helps assess the confidence and coherence of LLM explanations, offering insights into the LLMS' reasoning quality beyond accuracy alone.

\noindent \textbf{$\mathbf{RQ_4}$ -- What are the \underline{topics} most frequently mentioned by \acrshort{llm}s when answering questions related to the ROS2 software architectures?}
Identifying frequently mentioned topics reveals the focus and coverage of LLM reasoning in the context of ROS2 architectures, highlighting potential biases and gaps in architectural knowledge relevant to ROS2 systems.

\subsection{\acrshort{ros}2 Systems Selection}

To analyse the ability of \acrshort{llm}s to support the comprehension of the software architecture of \acrshort{ros}2 systems, we have selected three systems with computation graphs of different sizes. 
A description of the selected systems is provided below and in \autoref{tab:systemss}:
    (i) \href{https://docs.ros.org/en/humble/Tutorials/Beginner-Client-Libraries/Writing-A-Simple-Py-Publisher-And-Subscriber.html}{\texttt{\textbf{pubsub}}}: a baseline example of a \acrshort{ros}2 system consisting of a publisher and a subscriber node that are connected through a single topic~\cite{PUBSUB};
    %
     (ii)  \href{https://emanual.robotis.com/docs/en/platform/turtlebot3/simulation/}{\texttt{\textbf{turtlebot}}}: a modular mobile robot platform widely used in \acrshort{ros}2 education and reseach~\cite{Santiago:2023, Kashyap:2025, Krishna:2024} that can be run in simulation using Gazebo and RViz~\cite{TurtleBot3};
    (iii) 
    \href{https://github.com/IFRA-Cranfield/ros2_RobotSimulation/tree/humble}{\texttt{\textbf{panda}}}: a manipulator system integrating Gazebo and MoveIt~2 frameworks for simulation and motion planning, developed by the IFRA Group at Cranfield University to provide ready-to-use \acrshort{ros}2 robot simulation packages~\cite{panda:2022}.

The systems analysed in this study consist of (i) an official \acrshort{ros}2 tutorial example (\ie \texttt{pubsub}) and (ii) two third-party robot platforms (\ie \texttt{turtlebot} and \texttt{panda}); we use those systems since (i) their software architecture is heterogenous in terms of topology and size and (ii) they are developed by third parties, better supporting the external validity of our study. In this study, simulations are used in place of physical robots for reproducibility. The \acrshort{ros}2 Humble distribution is used for the whole experiment, as it is the only \acrshort{ros}2 release officially supported by \texttt{turtlebot} at the time the experiments were conducted.


\vspace{-5mm}

\subsection{\acrshort{llm}s Selection}

The choice of LLMs is based on the best-performing \acrshort{llm}s according to the LMArena leaderboard~\cite{llm_arena}. Specifically, the top ten unique models from the Arena Overview as ranked on June 16th (see \autoref{tab:llms}). While \texttt{gpt-4.5-preview} was originally in the list, this model has been deprecated and removed by OpenAI and is thus excluded from the experiment. For all models, we used the default settings, meaning no model parameters were specified during prompting.

\vspace{-3mm}
\begin{table}[htbp!]
    \centering
    \caption{Overview of the selected LLMs}
    \label{tab:llms}
    \footnotesize
    \vspace{-3mm}
    \begin{tabular}{lllp{1.2cm}rr}
         \textbf{LLM name} & \textbf{Developer} & \textbf{Release date} & \textbf{Context} & \multicolumn{2}{l}{\textbf{Cost (\$/MTok)}} \\ 
         
         & & & & \textit{Input} & \textit{Output} \\ \hline

         \texttt{gemini-2.5-pro} & Google & Jun 5, 2025 & 128k tokens & \$2.50 & \$10.00 \\
         \rowcolor{vu-grey-50}\texttt{o3-2025} & OpenAI & Apr 16, 2025 & 128k tokens & \$2.00 & \$8.00 \\
         \texttt{chatgpt-4o} & OpenAI & Mar 26, 2025 & 8k tokens & \$5.00 & \$20.00 \\
         \rowcolor{vu-grey-50}\texttt{claude-opus-4} & Anthropic & May 14, 2025 & 100k tokens & \$15.00 & \$30.00 \\
         \texttt{gemini-2.5-flash} & Google & May 20, 2025 & 64k tokens & \$1.00 & \$4.00 \\
         \rowcolor{vu-grey-50}\texttt{gpt-4.1} & OpenAI & Apr 14, 2025 & 8k tokens & \$4.00 & \$16.00 \\
         \texttt{grok-3} & xAI & Feb 24, 2025 & 16k tokens & \$3.00 & \$12.00 \\
         \rowcolor{vu-grey-50}\texttt{claude-sonnet-4} & Anthropic & May 14, 2025 & 100k tokens & \$3.00 & \$12.00 \\
         \texttt{o4-mini} & OpenAI & Apr 16, 2025 & 4k tokens & \$1.00 & \$4.00 \\ \hline
    \end{tabular}
\end{table}

\vspace{-5mm}

\subsection{JSON Encoding of \acrshort{ros} Computation Graphs}\label{sec:json}

Computation graphs are commonly used to reason about \acrshort{ros} systems. Existing tools can extract this information, for example, from bag files~\cite{Chen:2023} or in real time during system execution using the \texttt{rqt\_graph} package. In our study, we convert the computation graph to JSON following the structure proposed by Maruyama \etal~\cite{maruyama,ros2_performance_topology}. 
A JSON file is generated for each \acrshort{ros}2 system. 

\subsection{Prompt Generation and Evaluation}

Central to the experiment is the prompt plan, a general outline for the formulation of the prompts. \autoref{fig:prompt} explains the structure of the prompts used in this experiment. To limit the influence of the prompt on the response of the \acrshort{llm}s, the \textit{prompt} is treated as a controlled variable. However, no two generated prompts are identical as there are three placeholders in the prompt for the JSON topology, answer instruction, and question, respectively.


\subsection{Experimental Variables and Design}

The \textbf{independent variables} of this study are: (i) the \textit{\acrshort{llm}s} and (ii) the \acrshort{ros}2 \textit{systems}. The \textbf{dependent variables} are the: (i) \textit{number of output tokens}, (ii) \textit{accuracy}, (iii) \textit{perplexity}, and (iv) the topics mentioned in the contents of the \acrshort{llm}s' response. Accuracy is determined by comparing each answer with a predefined ground truth, yielding a binary evaluation of `correct' or `incorrect'. The accuracy for each combination of \acrshort{llm} and \acrshort{ros}2 system is calculated as the ratio of correct answers to the total number of questions. The number of output tokens is directly provided by the LLMs' APIs and it is meant to measure the verbosity of LLMs' responses. Topics and perplexity are described in Section~\ref{sec:data_analysis}.

The experiment follows a repeated measures design, in which each \acrshort{llm} answers the same sampled set of questions for each \acrshort{ros}2 system, allowing direct comparison of the performance of the models.
Each question is prompted to the \acrshort{llm}s in an identical way, following the prompt template in Listing~\ref{fig:prompt}. The prompt contains three placeholders corresponding to (i) the system topology, (ii) the answer instruction, and (iii) the question itself. The specific prompt used in this study was established and refined through several iterations. 
Initial tests were carried out to compare different ways of representing the \acrshort{ros}2 systems. The best results were achieved using the JSON topology, as \acrshort{llm}s are fundamentally textual models with a fixed context window. 
Trial runs were used to determine the formulation of the different types of questions and to specify the system's instructions. Finally, the Chain-of-Thought instruction was added to encourage the \acrshort{llm}s to reason step by step, improving the clarity and reliability of their responses.
We ultimately conducted the experiment by executing 1,230 prompts to each LLM, resulting in a total dataset of 1,230 x 9 = 11,070 prompts (and corresponding answers). 



{\captionsetup[figure]{name=Listing}
\begin{figure}[t]
    \centering
    \caption{Prompt template used in this study.} 
    \footnotesize
    \begin{boxx}
    \# System Prompt 

    The Robot Operating System (ROS) is a set of software libraries and tools for building robot applications. ROS2 is a middleware framework built on the Data Distribution Service (DDS) protocol.
    
    You are an AI assistant specialising in ROS2 robotic systems. 
    You can analyse and reason about robotic systems. 
    You aid ROS2 architects by answering questions about a given ROS2 system.

    ROS2 robotic systems are presented as system topologies in JSON format, including references to the ROS2 entities: nodes, topics, and services. 
    Communication between nodes is achieved through the anonymous publish/subscribe system or through the request/response mechanism between clients and service servers.
    
    Answer questions solely based on the explicit content of the input. 
    Do not infer, guess, or assume any information that is not present in the data. 
    If the question lacks sufficient context, state what additional information is required to answer the question effectively. 
    Respond by providing the direct answer to the question and a reference to where this can be found in the context provided.

    Be honest about the limitations of the data. Clarify uncertainty respectfully and avoid misleading conclusions.  \\
    
    \# Preamble 
    
    The following is a JSON topology of a ROS2 robotic system. 
    It lists the nodes, their publishers, subscribers, clients and service servers. \\
    
    \# Instruction 
    
    Analyse the <json> topology and answer the <question> strictly based on the provided <json> topology.

    <json>\{json\}<json> 
    
    Please provide your answer between the <answer></answer> tags and the explanation between the <explanation></explanation> tags.
    Your output must include \{answer instruction\}, and must not exceed the limit of 100 words.
    
    <question>\{question\}<question>

    Let's think step by step to be sure we have the right answer.
    \end{boxx}
    \label{fig:prompt}
\end{figure}}
    
\subsection{Experiment Execution}\label{sec:execution}

%
Before running the experiment, the topologies of the three \acrshort{ros}2 systems were retrieved in isolation. We opted to follow the guidelines as described by the maintainers for the simulations, specifically:

\begin{enumerate}[leftmargin=.5cm]
    \item \texttt{\textbf{pubsub:}} Create a package and include the code for the \texttt{talker} and \texttt{listener} nodes. Build the system and start the \texttt{talker} node. Then, start the \texttt{listener} node. Wait 10s before running the topology generation script. 
    \item \texttt{\textbf{turtlebot:}} Follow the `Quick Start Guide' to configure TurtleBot3 for \acrshort{ros}2 Humble. Build the system. Install the simulation package. Launch the simulation using an empty world. Start RViz2. Wait 10s before running the topology generation script. 
    \item \texttt{\textbf{panda:}} Install the project following the steps for the \acrshort{ros}2 Humble Environment. Import and install the improved \texttt{move\_group\_} \texttt{interface.h} file. Build the system. Launch Gazebo + MoveIt!2 Environment + ROS2 Robot Triggers/Actions. Select the empty world. Wait 10s before running the topology generation script. 
\end{enumerate}

Based on these topologies, questions were generated independently for each system according to \autoref{alg:questions}. The prompt plan (see \autoref{fig:prompt}) then sampled these questions following \autoref{eq:sample_size} and generated a structured JSON object to query the models.

The experiment was carried out between 30 June and 16 July, 2025 using APIs provided by the organisations. Each question was prompted one at a time for each \acrshort{llm}, and each question was presented only once. A script was used to run the experiment, iterating over the \acrshort{llm}s defined in a configuration file containing the specific settings of each \acrshort{llm} (\eg organisation, model name, batch support). The API requests included a specification of the specific model and the prompt. No configuration parameters were provided. The models' responses were logged as .TXT files, and the extracted message text was exported to a CSV file. 


\subsection{Data Analysis}\label{sec:data_analysis}

We extract the \textbf{number of tokens} used by each \acrshort{llm} from log files and then we compute descriptive statistics for models and systems. 

For assessing LLMs' \textbf{accuracy}, their answer within the \texttt{<answer>} tag is automatically extracted using a regex and compared to the ground truth. Comparisons are performed automatically on the basis of the general patterns identified. In total, 300 answers were labelled `incorrect' and we manually reviewed all of them to detect edge cases (\eg the closing \texttt{</answer>} tag was missing in 2 answers), resulting in the identification of  four additional `correct' answers.
We also sampled 273 correct answers, following a 95\% confidence interval and a 5\% error rate, and manually checked their correctness. 
For each combination of the \acrshort{llm} and \acrshort{ros}2 system, the response accuracy is determined by dividing the number of correct answers by the total number of questions for the respective system. 
Moreover, the computation graphs of the systems are plotted to depict the nodes and their connections through topics and services. 
The graphs are coloured according to the number of incorrect answers made concerning the specific entities.

\textbf{Perplexity} is used as a complementary metric to evaluate the quality of the explanations generated by the \acrshort{llm}s. 
The perplexity of the models' responses is used to measure the certainty of the \acrshort{llm}s. Perplexity of an answer $a$ is defined as
$
\text{Perplexity}(a) = 2^{\left( - \frac{1}{M} \sum_{j=1}^{M} \log P(t_j \mid t_{<j}, p) \right)}
$, where: 

\vspace{-2mm}
\begin{itemize}
    \item $a = (t_1, t_2, \dots, t_M)$ is the answer produced by the LLM;
    \item $M$ = number of tokens in the answer;
    \item $t_j$ = the $j$-th token in the answer;
    \item $t_{<j}$ = all tokens before $t_j$ in the answer;
    \item $p$ = the prompt provided to the LLM;
    \item $P(t_j \mid t_{<j}, p)$ = probability assigned by the LLM to token $t_j$ given the prompt and previous tokens.
\end{itemize}
\vspace{-1mm}
We compute $P$ by using a pre-trained GPT-2 model and tokenizer since (i) it is efficient in terms of time and computational resources, (ii) its logits and probabilities for each token are already available via the Hugging Face Transformers library~\cite{transformers}, (iii) it provides a consistent tokenization for all outputs of all 9 LLMs used in this study. 
Intuitively, perplexity indicates how predictable a sequence produced by a model (\ie one of the 9 selected LLMs) is to the model computing it (GPT‑2 in our case).
Lower perplexity values indicate that the model generates more coherent explanations, reflecting higher internal consistency in the provided answer.

We used \textit{topic modelling}~\cite{Hanker:2025} for identifying \textbf{mentioned topics} in the models' reasoning across responses. The analysis focusses exclusively on the text enclosed within the \texttt{<explanation>} tags. Before modelling, entity names and question-related terminology (\eg `topology', `node', `topic') are removed. The remaining text is subsequently cleaned, tokenised, and lemmatised before applying BERTopic~\cite{Grootendorst:2022}. 
BERTopic was selected due to its ability to extract novel insights from short and unstructured text and consistent performance in different domains~\cite{Egger:2022}.
A manual examination of a sample of 234 responses revealed recurring patterns about the use of list structures, references to the topology, and the inclusion of instructional cues. To quantitatively assess the prevalence of these patterns, a set of regular expressions was applied to systematically identify and count their occurrences within the results. 


\section{Questions Generation}

To ensure coverage of the various architectural components of the \acrshort{ros} systems, six categories of questions were established that cover entities in general, the different types of interface, and the connection between nodes. The type of questions can be multiple-choice (MCQ), boolean (BOOL), or open-ended (OPEN). 

\begin{algorithm*}[htbp!]
\caption{Systematic generation of questions about a ROS2 computation graph}\label{alg:questions}
    \footnotesize
    \KwInput{R = ROS2 computation graph}
    \KwOutput{Q = \{ Level, Category, Type, Question, Answer \}}
    
    \Begin{
        $N \gets$ ROS2 nodes defined in $R$; $S \gets$ ROS2 services defined in $R$; $T \gets$ ROS2 topics defined in $R$\;
        \ForEach {$e \in R_f$} {
            $Q \gets Q \cup \{(
                0,
                \text{ENTITY},
                \text{BOOL},
                \text{``Is there a ROS2 entity called <e>?''},
                \text{``No''}
            )\}$\;
        }
        \ForEach {$e \in R$} {
            $Q \gets Q \cup \{(
                0,
                \text{ENTITY},
                \text{BOOL},
                \text{``Is there a ROS2 entity called <e>?''},
                \text{``Yes''}
            )\}$\;
            $Q \gets Q \cup \{(
                0,
                \text{ENTITY},
                \text{MCQ},
                \text{``What kind of ROS2 entity is <e>? Possible answers: 1- a ROS topic, 2- a ROS service,} \newline \text{3- a ROS node.''}, 
                \text{\textit{find\_answer()}}
            )\}$\;
        }
        \ForEach {$n \in N$} {
            \ForEach {$t \in T$} {
                $Q \gets Q \cup \{(
                    1,
                    \text{PUBLISH},
                    \text{BOOL},
                    \text{``Does node <n> publish to topic <t>?''}, 
                    \text{\textit{find\_answer()}}
                )\}$\;
                $Q \gets Q \cup \{(
                    1,
                    \text{SUBSCRIBE},
                    \text{BOOL},
                    \text{``Is node <n> subscribed to topic <t>?''}, 
                    \text{\textit{find\_answer()}}
                )\}$\;
            }
            $Q \gets Q \cup \{(
                1,
                \text{PUBLISH},
                \text{OPEN},
                \text{``To which topics can node <n> publish?''}, 
                \text{\textit{find\_answer()}}
            )\}$\;
            $Q \gets Q \cup \{(
                1,
                \text{SUBSCRIBE},
                \text{OPEN},
                \text{``To which topics is node <n> subscribed?''}, 
                \text{\textit{find\_answer()}}
            )\}$\;
            \ForEach {$s \in S$} {
                $Q \gets Q \cup \{(
                    1,
                    \text{SERVICE},
                    \text{BOOL},
                    \text{``Does node <n> provide service <s>?''}, 
                    \text{\textit{find\_answer()}}
                )\}$\;
                $Q \gets Q \cup \{(
                    1,
                    \text{CLIENT},
                    \text{BOOL},
                    \text{``Does node <n> use service <s> as a client?''}, 
                    \text{\textit{find\_answer()}}
                )\}$\; 
            }
            $Q \gets Q \cup \{(
                1,
                \text{SERVICE},
                \text{OPEN},
                \text{``Which services does node <n> provide?''}, 
                \text{\textit{find\_answer()}}
            )\}$\;
            $Q \gets Q \cup \{(
                1,
                \text{CLIENT},
                \text{OPEN},
                \text{``Which services does node <n> use as a client?''}, 
                \text{\textit{find\_answer()}}
            )\}$\;
            \ForEach {$n_i \in N$} {
                \If {$n_i != n$} {
                    $Q \gets Q \cup \{( 
                        2,
                        \text{MESSAGE},
                        \text{BOOL},
                        \text{``Is there a communication path from node <n> to node <$n_i$> via a topic or service?''}, \newline
                        \text{\textit{find\_answer()}}
                    )\}$\;
                }
            }
        }
        \ForEach {$s \in S$} {
            $Q \gets Q \cup \{(
                1,
                \text{SERVICE\_TYPE},
                \text{OPEN},
                \text{``What is the type of service <s>?''}, 
                \text{\textit{find\_answer()}}
            )\}$\;
        }
        \ForEach {$t \in T$} {
            $Q \gets Q \cup \{(
                1,
                \text{TOPIC\_TYPE},
                \text{OPEN},
                \text{``What is the type of topic <t>?''}, 
                \text{\textit{find\_answer()}}
            )\}$\;
        }
    }
    
\end{algorithm*}

Based on the JSON-encoding of a ROS2 computation graph (see Section~\ref{sec:json}), a set of questions is generated according to Algorithm~\ref{alg:questions}. First all entities (\ie unique nodes, topics, and service servers) are extracted from the topology. For each entity, a question is generated to determine whether it is an entity or not, and a question on what kind of entity it is. Moreover, fake entities names are generated based on the different parts of actual entity names. The number of fake entities generated is equal to the number of real entities in the system. For these questions, only a question is included as to whether it is an entity in the \acrshort{ros} system. Then, for each node, boolean questions are generated for each possible topic or service to cheque whether the node publishes/subscribes to the topic or provides the service or is client to the service server. There is also an open question for the publishers, subscribers, services, and clients of each node. In addition, for every combination of two nodes, a boolean question is included on whether there is a communication path from one to the other. Lastly, for each service server and topic, an open question is generated on what the type is. 

The total number of questions generated for each system $s$ is $|Q_s| \approx |N_s| \times (|N_s| + 3|S_s| + 3|T_s| + 5) + 2|S_s| + 2|T_s| + |R_{f,s}|$ (where $N_s$ the set of all nodes in $s$, $S_s$ the set of all services in $s$, $T_s$ the set of all topics in $s$, and $R_{f,s}$ the set of false entities generated for $s$). For example, for the \texttt{pubsub} system (with $|N_{ps}|=2$, $|S_{ps}|=12$, and $|T_{ps}|=3$ ), this results in $|Q_{ps}| \approx 2 \times (2 + 3 \times 12 + 3 \times 3 + 5) + 2 \times 12 + 2 \times 3 + 2 = 136$ generated questions. 

\vspace{-3mm}
\begin{equation}
    S = 
    \begin{cases}
    30, & \text{if } \left\lfloor 0.1 \times G \right\rfloor < 30 \\
    \left\lfloor 0.1 \times G \right\rfloor, & \text{if } 30 \leq \left\lfloor 0.1 \times G \right\rfloor \leq 100 \\
    100, & \text{if } \left\lfloor 0.1 \times G \right\rfloor > 100
    \end{cases}
\label{eq:sample_size}
\end{equation}
\vspace{-3mm}

Since the number of questions generated is in $\Omega(N^2)$, we opted to sample the questions according to \autoref{eq:sample_size}. The sample size $S$ was determined for each category and question type, where $G$ represents the total number of generated questions. A minimum of 30 questions were sampled when available; otherwise, all questions were included. In general, we selected 10\% of the questions generated, with an upper limit of 100. \autoref{tab:sample} shows the population and corresponding sample sizes for each type of question per system. 

\begin{table*}[htbp]
    \centering
    \caption{Overview of the number of questions generated for each system, with sample sizes given in (parentheses).} 
    \footnotesize
    \vspace{-3mm}
    \begin{tabular}{r||c|c|c|c|c|c|c|c|c|c|c|c|c||c}
         \textit{Category} & \multicolumn{2}{c|}{\textit{Entity}} & \multicolumn{2}{c|}{\textit{Publish}} & \multicolumn{2}{c|}{\textit{Subscribe}} & \multicolumn{2}{c|}{\textit{Service}} & \multicolumn{2}{c|}{\textit{Client}} & \textit{Message} & \textit{S\_Type} & \textit{T\_Type} & \\ 
         
         {Type} & {BOOL} & {MCQ} & {BOOL} & {OPEN} & {BOOL} & {OPEN} & {BOOL} & {OPEN} & {BOOL} & {OPEN} & {BOOL} & {OPEN} & {OPEN} & {\textbf{TOTAL}} \\ \hline
         
         \texttt{\textbf{pubsub}} & 34 & 17 & 6 & 2 & 6 & 2 & 24 & 2 & 24 & 2 & 2 & 12 & 3 & \textbf{136} \\
            & (30) & (17) & (6) & (2) & (6) & (2) & (24) & (2) & (24) & (2) & (2) & (12) & (3) & \textbf{(132)} \\ 

         \rowcolor{vu-grey-50}
         \texttt{\textbf{turtlebot}} & 152 & 76 & 120 & 8 & 120 & 8 & 424 & 8 & 424 & 8 & 58 & 53 & 15 & \textbf{1,472} \\
            \rowcolor{vu-grey-50}
             & (30) & (30) & (30) & (8) & (30) & (8) & (43) & (8) & (43) & (8) & (30) & (30) & (15) & \textbf{(313)} \\ 
         
         \texttt{\textbf{panda}} & 698 & 349 & 1,320 & 40 & 1,320 & 40 & 11,040 & 40 & 11,040 & 40 & 1,560 & 276 & 33 & \textbf{27,796} \\
             & (70) & (35) & (100) & (30) & (100) & (30) & (100) & (30) & (100) & (30) & (100) & (30) & (30) & \textbf{(785)} \\ \hline

        \textbf{TOTAL} & 884 & 442 & 1,446 & 50 & 1,446 & 50 & 11,488 & 50 & 11,488 & 50 & 1,620 & 341 & 51 & \textbf{29,406} \\
            \rowcolor{vu-grey-50} 
             & (130) & (82) & (136) & (40) & (136) & (40) & (167) & (40) & (167) & (40) & (132) & (72) & (48) & \textbf{(1,230)} \\ \hline
    \end{tabular}
    \label{tab:sample}
\end{table*}
\section{Results}\label{s:results}

\subsection{Output Size ($\mathbf{RQ_1}$)}

\begin{figure*}[htbp]
    \centering
    \includegraphics[width=0.95\linewidth]{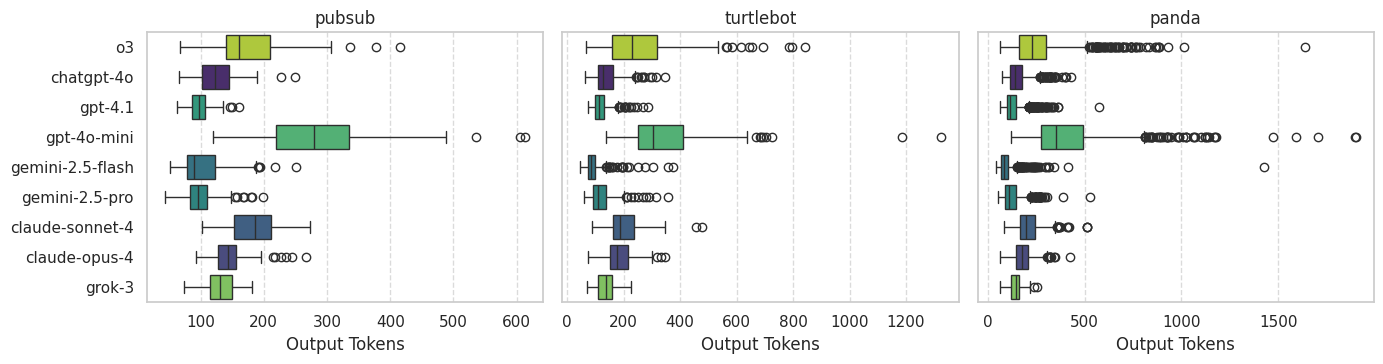}
    \vspace{-4mm}
    \caption{Distribution of the output tokens used by the \acrshort{llm}s across the \acrshort{ros}2 systems.}
    \label{fig:output_tokens}
\end{figure*}

An overview of the number of output tokens used per \acrshort{llm} and \acrshort{ros}2 system is shown in \autoref{fig:output_tokens}. In general, the number of tokens used increases with the complexity of the systems with an overall mean of 150, 177, and 293 for the \texttt{pubsub}, \texttt{turtlebot}, and \texttt{panda}, respectively. Only in the case of \texttt{gemini-2.5-flash} is the average number of tokens used to answer the questions related to \texttt{pubsub} the highest. In all other cases, an increase in system complexity results in an increase in tokens used by the models. The responses of \texttt{gpt-4o} were generally longer with on average 386 tokens, and more dispersed with a standard deviation is 195.

\begin{figure*}[htbp]
    \centering
    \includegraphics[width=0.95\linewidth]{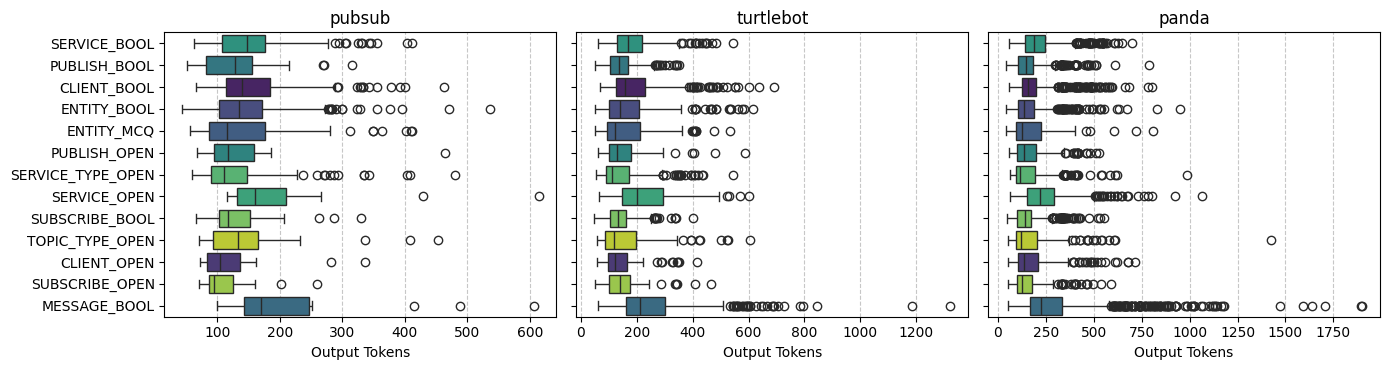}
    \vspace{-4mm}
    \caption{Distribution of the output tokens used for the different questions across the \acrshort{ros}2 systems.}
    \label{fig:question_tokens}
\end{figure*} 

\begin{figure}[htbp]
    \centering
    \includegraphics[width=0.95\linewidth]{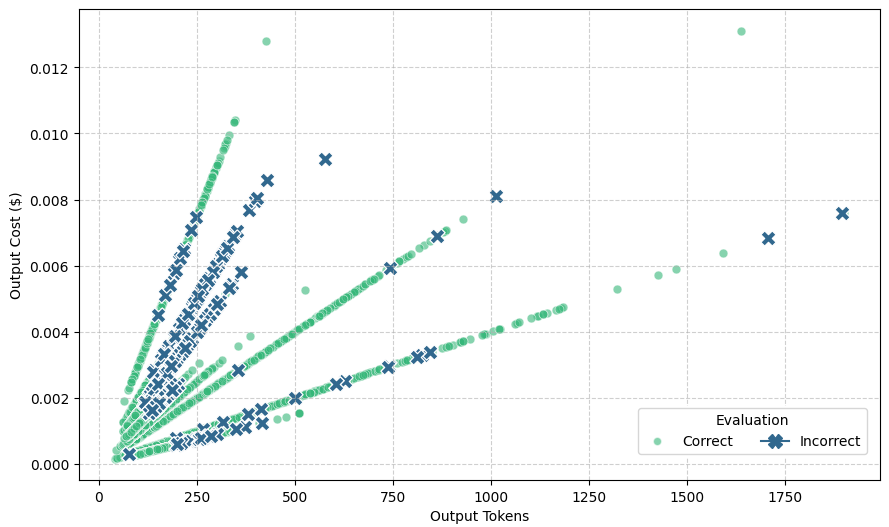}
    \vspace{-4mm}
    \caption{Number of output tokens against the costs.}
    \label{fig:output}
\end{figure}

The distribution of the number of tokens used per type of question (\ie category and type) shows that especially the MESSAGE\_BOOL questions result in a large number of tokens used (see \autoref{fig:question_tokens}). The tokens necessary also increases with the size of the system, from a mean of 226 for \texttt{pubsub} to on average 273 for \texttt{turtlebot} and 308 for \texttt{panda}. \autoref{fig:output} shows the number of tokens used for each response compared to the cost of the responses related to the distribution of correct and incorrect responses.

\begin{answerbox}
\textbf{Answer to $\mathbf{RQ_1}$} -- The number of output tokens increases with the complexity and size of the architecture of the system. \texttt{o3} and \texttt{gpt-4o-mini} tend to have higher variance across all systems, with the latter generating more tokens than others.
\end{answerbox}

\subsection{Accuracy ($\mathbf{RQ_2}$)}

The vast majority of questions were answered correctly (97.9\%), with even a 100\% accuracy on the \texttt{pubsub} questions across all \acrshort{llm}s. In total, 300 questions were incorrectly answered. There were 51 incorrect answers related to \texttt{turtlebot}, and 249 regarding \texttt{panda}. More than 98\% (297/300) of the incorrect answers refer to the MESSAGE\_BOOL questions. Of the 132 questions total in this category, 108 were incorrectly answered at least once. The other three incorrect answers were concerning the CLIENT\_OPEN questions by \texttt{chatgpt-4o} and \texttt{o4-mini}, and SUBSCRIBE\_OPEN by \texttt{gemini-2.5-flash}, all in the \texttt{panda} system. Table~\ref{tab:accuracy} shows the accuracy of the \acrshort{llm}s per \acrshort{ros} system. 

Analysis of the incorrect answers reveals that both \texttt{gemini-2.5-flash} and \texttt{o4-mini} extracted topics and services that were not in the topology for the SUBSCRIBE\_OPEN and CLIENT\_OPEN questions respectively. In the case of \texttt{gemini-2.5-flash} the correct topics were included in the answer; however, an additional topic was added to the list. \texttt{o4-mini} included a service while none was listed in the topology. \texttt{chatgpt-4o}, on the other hand, extracted the correct list of clients but excluded several on purpose, resulting in an incomplete list in the answer. The \acrshort{llm} reasoned that, because ``we are only interested in services being used as clients'', only entries with the term ``srv'' would qualify, and the ones included ``action'' were therefore excluded from the final answer. 

\begin{table}[htbp]
    \centering
    \caption{Overall accuracies of the models per system.}
    \footnotesize
    \vspace{-3mm}
    \begin{tabular}{r|ccc|c}
         \textbf{LLM} & \textbf{\texttt{pubsub}} & \textbf{\texttt{turtlebot}} & \textbf{\texttt{panda}} & \textbf{Mean} \\ \hline
         
         \texttt{gemini-2.5-pro} 
            & 100.0\% 
            & 100.0\% 
            & 100.0\% 
            & 100.0\%\\ 
         
         \rowcolor{vu-grey-50}\texttt{o3} 
            & 100.0\% 
            & 99.68\%  
            & 99.62\% 
            & 99.77\% \\ 
            
         \texttt{chatgpt-4o} 
            & 100.0\% 
            & 96.49\%  
            & 91.34\% 
            & 95.94\% \\ 
            
         \rowcolor{vu-grey-50}\texttt{claude-opus-4} 
            & 100.0\% 
            & 99.04\%  
            & 98.98\% 
            & 99.34\% \\  
            
         \texttt{gemini-2.5-flash} 
            & 100.0\% 
            & 99.68\%  
            & 99.49\%
            & 99.72\% \\ 
            
         \rowcolor{vu-grey-50}\texttt{gpt-4.1} 
            & 100.0\% 
            & 95.53\%  
            & 90.45\%
            & 95.33\% \\ 
            
         \texttt{grok-3} 
            & 100.0\% 
            & 94.89\%  
            & 91.97\% 
            & 95.62\% \\ 
            
         \rowcolor{vu-grey-50}\texttt{claude-sonnet-4} 
            & 100.0\% 
            & 99.04\%  
            & 97.83\% 
            & 98.96\% \\  
         
         \texttt{o4-mini} 
            & 100.0\% 
            & 99.36\%  
            & 98.60\%  
            & 99.32\% \\ \hline

        \textbf{Mean}
            & 100.0\%
            & 98.19\% 
            & 96.48\% 
            & 98.22\% \\
         
    \end{tabular}
    \label{tab:accuracy}
\end{table}

The \acrshort{llm}s mainly struggle with the MESSAGE\_BOOL questions, specifically the ones where the topic \texttt{'/parameter\_events'} provides the sole communication path. This behaviour is expected, since all nodes publish by default to this topic. The dataset used in the study included a total of 963 such questions (107 unique), and 293 times these questions were incorrectly answered. 
In the cases where it was not considered a valid path, more than half of the time the analysis was incomplete (152, 52.1\%) because not all interfaces were identified. However, when \texttt{'/parameter\_events'} was correctly listed as one of the interfaces, the connection was often ignored (63, 21.6\%) or identified as an invalid path (58, 19.9\%) because it is a `global', `system-wide' topic that `does not constitute a unique communication path'. Various times the \acrshort{llm}s mixed up the list of subscribers and publishers, resulting in a wrong conclusion (16, 5.5\%). Three times the question was answered in the wrong direction (1.0\%). 
In the other 5 cases, the \acrshort{llm}s identified a communication path, while this was not the case; \texttt{claude-sonnet-4} made a directionality error twice and once by \texttt{claude-opus-4}, \texttt{gemini-2.5-flash} extracted a list of subscribers that did not exist, and \texttt{claude-opus-4} confused a correctly retrieved subscriber for a publisher, resulting in a non-existent path. 

\begin{figure}[h!]
    \centering
    \includegraphics[width=0.95\linewidth]{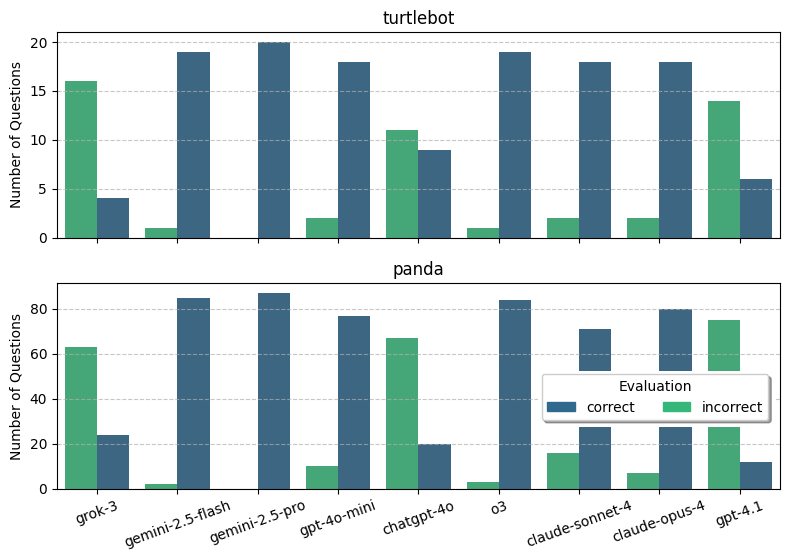}
    \vspace{-4mm}
    \caption{LLMs' accuracy when the only possible communication path is through \texttt{'/parameter\_events'}.}
    \label{fig:parameterevents}
\end{figure}

In total, each model received 107 MESSAGE\_BOOL questions where the only possible path is through \texttt{'/parameter\_events'}. While \texttt{gemini-2.5-pro} consistently answered these questions positively, all other models varied in their analysis (see \autoref{fig:parameterevents}). \texttt{gpt-4.1}, \texttt{grok-3}, and \texttt{chatgpt-4o} more often said there was not a communication path, 89, 79, and 78 times, respectively. The other \acrshort{llm}s did analyse a communication path at least 83.2\% of the times. In 15 of the cases where the \acrshort{llm} answered positively, the response did not include a reference to the topic \texttt{'/parameter\_events'}. 

There are 14 cases where the \acrshort{llm} stated that it could not determine whether a node used a specific service or client because the client information was missing. Based on this, \acrshort{llm}s responded negatively to the possibility of a connection, and the responses provided were evaluated as \textit{correct} since the missing field indicates that the node is not a client for any service server.

The analysis of the computation graphs (see our replication package~\cite{replication_package}) tell us that for \texttt{pubsub} all nodes and topics are correctly analysed by the LLMs, whereas for  \texttt{turtlebot} and \texttt{panda} we observe at least one incorrect answer for 25.0\% and 22.5\% of the nodes, respectively. If we do not consider the \texttt{'/parameter\_events'} topic, the majority of the incorrect answers is related to a very low number of nodes within the 3 studied systems, bringing the maximum number of incorrect answers per node down to only two. 

\begin{answerbox}
\textbf{Answer to $\mathbf{RQ_2}$} -- Accuracy is high across LLMs (mean=98.22\%), with perfect accuracy on the simplest system (\ie \texttt{pubsub}). \texttt{gemini-2.5-pro} performs best (100\%), followed by \texttt{o3} (99.77\%) and \texttt{gemini-2.5-flash} (99.72\%), while \texttt{gpt-4.1} shows the lowest accuracy (95.33\%). Of the 11,070 prompts executed in this study, only 300 are answered incorrectly, mostly involving the largest system (\ie \texttt{panda}) and a single question type (message exchange detection involving parameter changes).
\end{answerbox}

\subsection{Perplexity ($\mathbf{RQ_3}$)}\label{sec:perplexity}

There is a wide disparity in the mean perplexity across LLMs. \texttt{o4-mini} has the highest score with 103.6, while the lowest is achieved by \texttt{chatgpt-4o} with 19.6. The Anthropic models achieve similar scores with \texttt{claude-sonnet-4} 20.6 and \texttt{claude-opus-4} 20.0. The Google models resulted in a score of 33.5 for \texttt{gemini-2.5-pro} and 48.7 for \texttt{gemini-2.5-flash}, and xAI's \texttt{grok-3} score is 35.5. The mean perplexity for \texttt{o3} is 70.4 and for \texttt{gpt-4.1} it is 27.5.

A common pattern identified in the explanations of the \acrshort{llm}s is the usage of lists. Numbered lists were used 472 times by different models, and more than 25\% of the responses included dotted or dashed lists (2,849/11,070). Especially \texttt{claude-opus-4} and \texttt{claude-sonnet-4} used both types of lists often. In contrast, \texttt{grok-3} did not include any type of list in its response. In addition, all models occasionally reference the message or service types in their responses to questions other than SERVICE\_TYPE\_OPEN and MESSAGE\_TYPE\_OPEN. Some models include an explicit instruction that they ``need to check'' in their response, in the case of \texttt{claude-sonnet-4} the instruction is included in 52.2\% of the responses. 

\begin{answerbox}
\textbf{Answer to $\mathbf{RQ_3}$} -- The mean perplexity exhibited by LLMs varies substantially, from 19.6 (\texttt{chatgpt-4o}) to 103.6 (\texttt{o4-mini}), indicating notable differences in the stability and predictability of the generated explanations for ROS2 software architecture-related answers. 
\end{answerbox}

\subsection{Mentioned Topics ($\mathbf{RQ_4}$)}

In total, six distinct topics were identified in the explanations provided by the \acrshort{llm}s. \autoref{tab:topics} includes the specific words associated with each topic, the coherence of the topic, and the total number of occurrences in all responses. 

\vspace{-3mm}
\begin{table}[htbp!] 
    \centering
    \caption{Topics identified in the explanations of the \acrshort{llm}s.}
    \footnotesize
    \vspace{-3mm}
    \begin{tabular}{lp{1.4cm}p{3.8cm}rr}
         & \textbf{Topic} & \textbf{Words} & \textbf{Coherence} & \textbf{Count} \\ \hline

        0 & \textit{service references} 
            & [`server', `service', `list', `provides', `provided', `client', `type', `array']
            & 0.394
            & 3,710 \\
            
        \rowcolor{vu-grey-50}1 & \textit{publisher references} 
            & [`publisher', `type', `publishes', `list', `message', `topic', `msg', `published']
            & 0.409
            & 2,019 \\
            
        2 & \textit{system entities} 
            & [`entity', `ro', `name', `server', `node', `type', `named', `provided']
            & 0.720
            & 1,793 \\
            
        \rowcolor{vu-grey-50}3 & \textit{subscriber references} 
            & [`subscriber', `subscribed', `list', `topic', `subscription', `msg', `type', `listed']
            & 0.406
            & 1,574 \\

        4 & \textit{communication path} 
            & [`communication', `path', `subscribes', `publishes', `via', `direct', `topic', `node']
            & 0.762
            & 1,182 \\
            
        \rowcolor{vu-grey-50}5 & \textit{interfaces} 
            & [`client', `server', `publisher', `subscriber', `listed', `service', `field', `us']
            & 0.536
            & 792 \\ \hline
         
    \end{tabular}
    \label{tab:topics}
\end{table}
\vspace{-2mm}

\autoref{tab:llm_topics} shows how the topics are distributed in the different \acrshort{llm}s. In general, the distribution is uniform to global counts. However, the topic \textit{service references} occur significantly less in the responses of \texttt{claude-opus-4} and \texttt{claude-sonet-4}, while the \textit{interfaces} are more present in the answers provided by those two LLMs.

\vspace{-3mm}
\begin{table}[htbp!]
    \centering
    \caption{Distribution of the topics per \acrshort{llm}.}
    \footnotesize
    \vspace{-3mm}
    \begin{tabular}{r|cccccc}

        \textbf{\acrshort{llm}} & \textbf{0} & \textbf{1} & \textbf{2} & \textbf{3} & \textbf{4} & \textbf{5}  \\ \hline
    
         \texttt{gemini-2.5-pro} & 405 & 224 & 208 & 175 & 132 & 86 \\
         
         \rowcolor{vu-grey-50}\texttt{o3-2025} & 433 & 218 & 195 & 178 & 131 & 75 \\
            
         \texttt{chatgpt-4o} & 435 & 224 & 213 & 177 & 132 & 49 \\
            
         \rowcolor{vu-grey-50}\texttt{claude-opus-4} & 337 & 241 & 207 & 156 & 132 & 157 \\
            
         \texttt{gemini-2.5-flash} & 447 & 220 & 192 & 180 & 130 & 61 \\
            
         \rowcolor{vu-grey-50}\texttt{gpt-4.1} & 434 & 225 & 211 & 172 & 133 & 55 \\
         
         \texttt{grok-3} & 420 & 226 & 206 & 177 & 133 & 68 \\
            
         \rowcolor{vu-grey-50}\texttt{claude-sonnet-4} & 339 & 224 & 211 & 175 & 132 & 149 \\
         
         \texttt{o4-mini} & 460 & 217 & 150 & 184 & 127 & 92 \\ \hline

    \end{tabular}
    \label{tab:llm_topics}
\end{table}

Although there are more categories than topics (see \autoref{tab:category_topics}), categories can be grouped into the identified topics. For \textit{service}, \textit{publish}, \textit{entities}, \textit{subscribe}, and \textit{message} questions, the mapping is one-to-one with the topics. In the other cases, categories are strongly related (\eg \textit{client} with \textit{service}) or a subcategory (\eg \textit{S\_Type} of \textit{Service}). The \textit{T\_Type} questions, however, are mostly amongst \textit{publisher references} topic while this category of questions is also closely related with the \textit{subscriber references}. 

\vspace{-3mm}
\begin{table}[htbp]
    \centering
    \caption{Distribution of the topics per category.}\label{tab:category_topics}
    \vspace{-4mm}
    {\footnotesize
    \begin{tabular}{l|cccccc}

        \textbf{Category} & \textbf{0} & \textbf{1} & \textbf{2} & \textbf{3} & \textbf{4} & \textbf{5}  \\ \hline
    
         \textit{Entity} & 123 & 18 & \textbf{1,753} & 4 & 2 & 8 \\

         \rowcolor{vu-grey-50}\textit{Publish} & 5 & \textbf{1,558} & 1 & 16 & 1 & 3 \\

         \textit{Subscribe} & 1 & 31 & 4 & \textbf{1,537} & 0 & 11 \\

         \rowcolor{vu-grey-50}\textit{Service} & \textbf{1,734} & 0 & 12 & 0 & 0 & 117 \\

         \textit{Client} & \textbf{1,208} & 0 & 11 & 0 & 0 & 644 \\

         \rowcolor{vu-grey-50}\textit{Message} & 0 & 0 & 0 & 1 & \textbf{1,179} & 8 \\
         
         \textit{S\_Type} & \textbf{638} & 0 & 9 & 0 & 0 & 1 \\

        \rowcolor{vu-grey-50}\textit{T\_type} & 1 & \textbf{412} & 3 & 16 & 0 & 0 \\ \hline

    \end{tabular}}
\end{table}
\vspace{-3mm}

\begin{answerbox}
\textbf{Answer to $\mathbf{RQ_4}$} -- Topic coherence ranges from 0.394 (\textit{service references}) to 0.762 (\textit{communication path}). 
The distribution of topics tends to be uniform across LLMs, with the only exception of Claude models which tend to mention more \textit{interfaces} and less \textit{service references}. 
\end{answerbox}

\section{Discussion}\label{s:discussion}



\noindent \textbf{General reflections}. Our results show that \acrshort{llm}s performed well as assistants for architecture comprehension tasks on ROS2 systems. All models provided  correct answers about the \texttt{pubsub} system. However, the accuracy of most models decreased as system complexity increased, particularly for \texttt{panda}.  
The majority of the mistakes made were related to communication paths, indicating that the models struggle with reasoning about implicit or indirect communication paths. Specifically, there was no consensus amongst the models regarding the purpose of the \texttt{'/parameter\_events'} topic. The variation in handling the system-level topic between models indicates differences in internal reasoning and training exposure to \acrshort{ros}2 conventions. Although \texttt{gemini-2.5-pro} consistently accepted it as a valid link, other models, including \texttt{gpt-4.1} and \texttt{chatgpt-4o}, tended to reject it. This variation indicates that \acrshort{llm}s differ in how they conceptualise global communication topics, which may affect their reliability in more complex reasoning tasks.
There were several occasions in which the models hallucinated. For example, \texttt{both gemini-2.5-flash} and \texttt{o4-mini} subtracted a list of topics or services that were not in the topology. In addition, some models occasionally hallucinated a communication path through \texttt{'/parameter\_events'}. However, this was also the result of interpreting the direction of the question incorrectly. 

The results of topic modelling highlight that \acrshort{llm}s tend to focus on enumerating structural entities rather than abstract relationships or behaviours. The strong presence of references in the explanations, along with frequent list-based reasoning, reinforces that the models are more confident in retrieval-based reasoning than in inferring implicit dependencies. Moreover, the presence of procedural markers such as ``need to check'' indicates that models often simulate reasoning through a stepwise instruction-like discourse. 

The mean perplexity scores reveal substantial differences in the models’ ability to generate confident responses to architecturally-relevant questions. \texttt{o4-mini} exhibits the highest perplexity, indicating that its output is the least predictable and potentially less fluent, while \texttt{chatgpt-4o} achieves the lowest score, suggesting more consistent and coherent responses. The differences suggest that perplexity can provide an additional lens for evaluating not only the correctness of responses but also the fluency and internal consistency of the explanations generated by each model. 

From a computational perspective, the increase in output tokens with system complexity reflects greater reasoning effort by the models. However, longer responses did not always correlate with higher accuracy. Moreover, \acrshort{llm}s that cost more tend to have shorter responses, but also slightly more incorrect answers. The dispersion in token usage, particularly by \texttt{chatgpt-4o}, also reflects variability in how models balance confidence and explanation detail.

\noindent \textbf{Implications for \acrshort{ros} developers}. Our results highlight both the potential and limitations of current \acrshort{llm}s as tools for supporting the comprehension of system topologies and supporting design validation. Although models can reliably extract explicit information from the topology, they show greater difficulty in reasoning about implicit or system-wide communication paths. Moreover, the results highlight the need for caution when prompting the models, as they have an inherent bias with respect to, for instance, system-level topics such as \texttt{/parameter\_events}. Based on ones own preferences, developers might prefer to \texttt{use gemini-2.5-pro}, for instance, over \texttt{gpt-4.1}, or vice versa. Our results also emphasise the need for further exploration of different prompting strategies, evaluate them using multiple \acrshort{ros} systems and \acrshort{llm}s.  
Further practical implications of our study include: 
(i) if cost is a top priority, \texttt{gemini-2.5-flash} is the LLM producing the lowest number of output tokens and it is the cheapest in terms of both input and output tokens, still with excellent results (see~\autoref{tab:accuracy});
(ii) if the system under analysis is highly complex and accuracy of the LLMs' answers is a top priority, then developers should use an LLM in the Gemini family, like \texttt{gemini-2.5-pro} (see \autoref{tab:accuracy});
(iii) developers should pay special attention to system-level entities like the \texttt{parameter\_events} topic when prompting LLMs since they might strongly influence the accuracy of the outcomes (see \autoref{fig:parameterevents});
(iv) we suggest to use multiple LLMs during a comprehension task since different LLMs provide answers with different reasoning structures and levels of perplexity, which can help architects getting a more nuanced understanding of the architecture of their ROS2 system.  

\noindent \textbf{Implications for researchers}. Our results point to promising avenues for fine-tuning or instruction alignment specifically targeted at improving reasoning tasks in the context of ROS2 systems. Specifically, building on the evidence provided in this work that LLMs perform reasonably well in comprehending basic facts about the software architecture of a ROS2 system, the software engineering community might explore the boundaries of such comprehension capabilities; for example, by answering the following research questions: To what extent can LLMs comprehend more complex facts about a ROS2 system (\eg presence of architectural antipatterns~\cite{mo2019architecture} or opportunities for refactoring~\cite{baqais2020automatic})? How can telemetry data be integrated into the architecture description prompted to the LLM (\eg via RAG or a tool-based AI agent) and used for extracting facts related to non-functional properties (\eg energy consumption~\cite{ICRA_2026},  performance~\cite{JBCS_2025})? How can LLMs be effectively integrated into a (mesh of) AI agents for supporting architecting activities~\cite{vaidhyanathan2025software}?
Another line of research with potential is the representation of \textit{architectural tactics}~\cite{bass2021software} into a knowledge based that LLMs can use to provide timely
recommendations about applicable tactics either at development time or at runtime, basically leading to LLM-supported architectural self-adaption for ROS2 systems~\cite{garlan2009software}.



\section{Threats To Validity}\label{sec:threats}

\noindent \textbf{Internal Validity}. 
We are aware about the LLMs' sensitivity to prompt phrasing. This concerns both the prompt plan and the formulation of the questions. 
To mitigate this risk, the prompt and questions were treated as a controlled variable: all models were evaluated using the same standardised prompts, which were iteratively refined until consistent responses were observed. 

\noindent \textbf{External Validity}. 
The chosen \acrshort{ros}2 systems might not represent the diversity of \acrshort{ros}2 deployments in real-world applications. To mitigate this threat, three systems of different sizes were considered to capture a range of system topologies and interface configurations, ensuring that the evaluation included both minimal setups and more elaborate architectures. 
All three systems were developed by third parties and were not created by the authors, which strengthens this study by ensuring that the evaluation reflects systems that could realistically be encountered in practice.
The choice of used \acrshort{llm}s form another threat to validity. However, the choice to only use the highest-rated models was made because the goal of this study is to evaluate current state-of-the-art capabilities in reasoning about \acrshort{ros}2 systems. In addition, the usage of nine models from four different providers strengthens the diversity of the evaluation. 


\noindent \textbf{Construct Validity}. 
Ambiguous or underspecified questions posed a potential threat to construct validity, as they could influence the quality and completeness of the model responses. 
To address this potential threat, multiple phrasings were iteratively tested and evaluated and a final version was selected that minimised ambiguity, while preserving the intended construct.
Moreover, the prompt included an instruction to use no more than 100 words in the response to limit the costs for this study. This restriction introduced a potential threat to validity, as it constrained the ability of some models to fully express their reasoning. 
%
The specific representation of the architecture of a \acrshort{ros}2 system used in this study represents a potential threat to validity. To mitigate this threat, several approaches were considered and compared. However, the best performance was achieved in the trial runs with the JSON topology, as included in the current work. 
The correctness of the LLMs' answers was assessed by automatically extracting the text enclosed in the \texttt{<answer>} tags and comparing it with a predefined ground truth using regexes. Although this approach ensures consistency and scalability, it also introduces a potential threat to the validity of the construct. Subtle differences in phrasing, formatting, or minor variations could be flagged as incorrect, potentially misinterpreting the answer of the model. To mitigate errors introduced by automated evaluation, all answers initially labelled as incorrect were manually reviewed. This step reduced the risk of misclassification due to edge cases, such as missing closing tags or minor formatting differences. By combining automated and manual checks, the evaluation aimed to accurately reflect the correctness of each response while maintaining the integrity of the construct being measured.
Another threat concerns the interpretation of the \acrshort{ros}2 topic \texttt{/parameter\_events}. This topic can be understood as solely a system-internal mechanism or also as a topic available for communication between nodes. Such ambiguity may lead \acrshort{llm}s to produce responses that are logically valid under one interpretation but appear incorrect under the evaluation criteria. To address this issue, responses concerning these questions were more thoroughly analysed to identify the models’ underlying motivations. Nevertheless, this issue illustrates how ambiguity in domain-specific concepts can affect whether the evaluation accurately measures the intended construct of reasoning about the architecture of \acrshort{ros}2 systems.

\noindent \textbf{Conclusion Validity}. 
Due to the differences in sizes between the considered systems, there was a wide disparity between the number of questions generated for each of the systems. To balance the questions per system and keep the experiment manageable, we decided to use a sampling strategy to select the questions used. For each system, question category and type, at least 30 and at most 100 questions were randomly sampled, if available. The minimum of 30 questions was chosen to leverage the Central Limit Theorem, which suggests that sample means tend to approximate a normal distribution for sample sizes of 30 or more, supporting more reliable statistical analysis. However, especially for the smaller \texttt{\textbf{pubsub}} system, there were often not 30 questions to sample from. In those cases, all questions were selected. However, smaller sample sizes can limit statistical power, increase variability, and reduce confidence in the conclusions drawn from the data. Although this approach ensured feasibility, it can affect the reliability of statistical comparisons between systems and models.

\section{Conclusion}\label{s:conclusion}


This study presents initial results on the ability of large language models (\acrshort{llm}s) to support the comprehension of \acrshort{ros}2 systems. We analysed the answers of nine \acrshort{llm}s to a total of 1,230 questions related to three different \acrshort{ros}2 systems of varying complexity. The responses were compared with a predefined ground truth to assess correctness, and topic modelling was applied to analyse the reasoning patterns.
The findings indicate that \acrshort{llm}s have the capability of supporting the comprehension of \acrshort{ros}2 concepts and can accurately answer architecturally-relevant questions. This suggests that they have potential to assist robotics architects, particularly in system exploration and validation tasks. However, differences between models and recurring misinterpretations of system-level communication reveal that their reasoning about structural relationships remains limited.

As future work, we will replicate this study using a broader range of (possibly industrial, large-scale) \acrshort{ros}2 systems and more recent \acrshort{llm}s to assess the generalisability of the findings. Additionally, the analysis will be extended to include other system entities, such as action servers, which serve as important interfaces for node interactions. Further investigations will explore the impact of prompt engineering techniques on LLMs' accuracy. 
Alternative representations of \acrshort{ros}2 architectures (\eg using \texttt{rospec}~\cite{canelas2025usability} will be used to evaluate the effect on the reasoning of the \acrshort{llm}s. 
We will also involve humans in verifying LLM-provided answers to strengthen the validity of the qualitative analysis and mitigate potential bias introduced by models' self-justification.
We also plan to ground LLM queries in ROS2 developers’ GitHub issues and pull request to have a deeper assessment of the capabilities of LLMs concerning ROS software architectures.

\begin{acks}
This work is supported by the InnoGuard Marie Skłodowska-Curie Doctoral Network (Grant Agreement No. 101169233).
\end{acks}

\bibliographystyle{ACM-Reference-Format}
\bibliography{references}


\end{document}